\documentclass[twocolumn,prb]{revtex4}
\usepackage{amsfonts}
\usepackage[T1]{fontenc}
\usepackage{amsmath,amsbsy,amssymb,graphicx}
\usepackage{times}

\begin{document}

\title{Nonlinear topological phase diagram in dimerized sine-Gordon model}
\author{Motohiko Ezawa}
\affiliation{Department of Applied Physics, University of Tokyo, Hongo 7-3-1, 113-8656,
Japan}

\begin{abstract}
We investigate the topological physics and the nonlinearity-induced trap
phenomenon in a coupled system of pendulums. It is described by the
dimerized sine-Gordon model, which is a combination of the sine-Gordon model
and the Su-Schrieffer-Heeger model. The initial swing angle of the left-end
pendulum may be regarded as the nonlinearity parameter. The topological
number is well defined as far as the pendulum is approximated by a harmonic
oscillator. The emergence of the topological edge state is clearly
observable in the topological phase by solving the quench dynamics starting
from the left-end pendulum. A phase diagram is constructed in the space of
the swing angle $\xi \pi $ with $|\xi |\leq 1$ and the dimerization
parameter $\lambda $ with $|\lambda |\leq 1$. It is found that the
topological phase boundary is rather insensitive to the swing angle for $%
|\xi |\lesssim 1/2$. On the other hand, the nonlinearity effect becomes
dominant for $|\xi |\gtrsim 1/2$, and eventually the system turns into the
nonlinearity-induced trap phase. Furthermore, when the system is almost
dimerized ($\lambda \simeq 1$), coupled standing waves appear and are
trapped to a few pendulums at the left-end, forming the dimer phase. Its
dynamical origin is the cooperation of the dimerization and the nonlinear
term.
\end{abstract}

\maketitle

\textit{Introduction}. Topological physics is extensively studied in
condensed-matter physics\cite{Hasan,Qi}. A topological phase is signaled by
the emergence of topological edge states if a sample has a boundary. Now,
the target of topological physics is expanded to artificial topological
systems such as acoustic\cite%
{Prodan,TopoAco,Berto,Xiao,He,Abba,Xue,Ni,Wei,Xue2}, mechanical\cite%
{Lubensky,Chen,Nash,Paul,Sus,Sss,Huber,Mee,Kariyado,Hannay,Po,Rock,Takahashi,Mat,Taka,Ghatak,Wakao}%
, photonic\cite{KhaniPhoto,Hafe2,Hafezi,WuHu,TopoPhoto,Ozawa,Hassan,Li} and
electric circuit\cite%
{TECNature,ComPhys,Hel,Lu,YLi,EzawaTEC,Zhao,EzawaLCR,EzawaSkin,Garcia,Hofmann}
systems. The merit of artificial topological systems is that system
parameters can be finely controlled. In addition, it is possible to make a
small sample with clean edges.

Topological physics has been mainly studied in linear systems. Recently, the
frontier of the study of topological phases has reached nonlinear systems.
Nonlinear effects are naturally introduced in artificial topological
systems. In this context, topological physics in nonlinear systems is
studied in mechanical\cite{Snee,PWLo,MechaRot}, photonic\cite%
{Ley,Zhou,MacZ,Smi,Tulo,Kruk,NLPhoto,Kirch}, electric circuit\cite%
{Hadad,TopoToda} and resonator\cite{Zange} systems. It is an interesting
problem how the topological phases are modified in the presence of the
nonlinear term. Furthermore, it is fascinating if there is a transition
induced by the nonlinear term.

\begin{figure}[b]
\centerline{\includegraphics[width=0.48\textwidth]{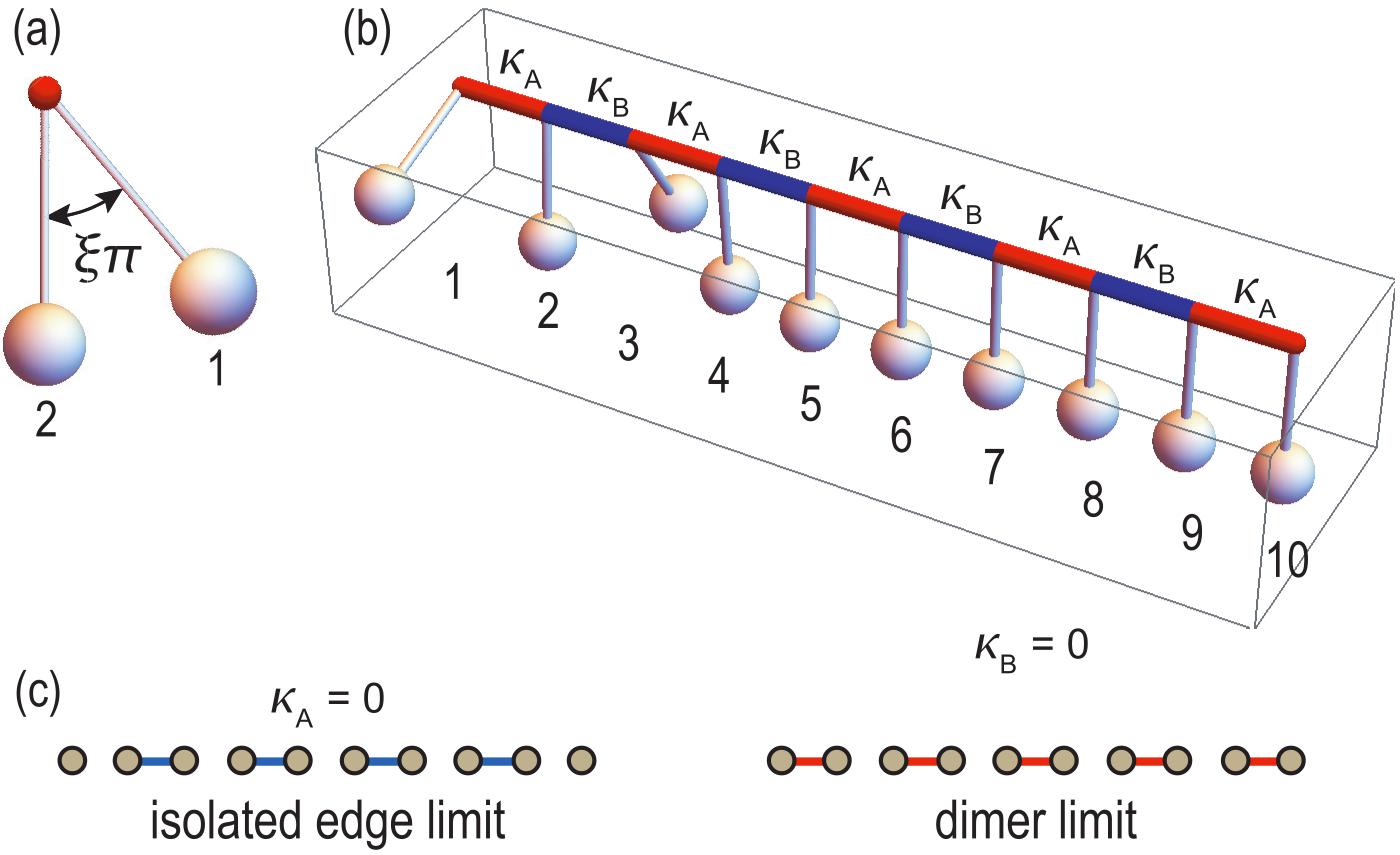}}
\caption{Illustration of coupled pendulums. Adjacent pendulums are connected
by a wire which has a restoring force depending on the angle difference. The
torsion is alternating as $\protect\kappa _{A}$ and $\protect\kappa _{B}$.
The system is described by the dimerized sine-Gordon equation. (a)
Horizontal view of the initial condition (\protect\ref{IniCon}). (b) Bird's
eye's view of a chain of pendulums in the topological phase, where $g/\protect\kappa =1,$ $\protect\lambda =-0.5$ and $\protect\xi =0.5$. 
The pendulum at $n=2$ is stationary.
(c) Illustration of the isolated edge limit and the dimer limit. }
\label{FigPendulum}
\end{figure}

In this paper, we study topological physics in nonlinear systems
analytically and numerically by taking the discrete sine-Gordon model with
dimerization. This model is realized by a coupled pendulum system with
alternating torsion as illustrated in Fig.\ref{FigPendulum}. We give an
oscillation only to the left-end pendulum initially, and investigate a
quench dynamics how the oscillation propagates to other pendulums. The
initial swing angle $\xi \pi $ at the left-end pendulum may be taken as the
nonlinearity parameter, where $|\xi |\leq 1$. We explore a phase diagram of
the model. Since this model is reduced to the Su-Schrieffer-Heeger (SSH)
model in the absence of the nonlinear term, it is natural to expect that the
topological phase diagram is valid in the weak nonlinear regime. Indeed, it
is possible to define the topological number for $|\xi |\ll 1$ by the
first-order perturbation theory with respect to $\xi $. Physically this
correspond to the case where the pendulum motion is well approximated by a
harmonic oscillator. Beyond the weak nonlinear regime we determine phases by
studying a quench dynamics numerically. It is found that the phase boundary
is quite insensitive to the nonlinearity parameter for $|\xi |\lesssim 1/2$.
The nonlinearity effect becomes dominant for $|\xi |\gtrsim 1/2$, and
eventually the system turns into the nonlinearity-induced trap phase.

We have found four phases. First, we have the topological phase, where the
oscillation at the left-end pendulum stays as a standing wave. Second, we
have the trivial phase, where it propagates into the bulk. Third, the system
turns into the trap phase in the strong nonlinear regime, where the
oscillation occurs as a perfectly localized standing wave at the left-end
pendulum. This is because the interaction between adjacent pendulum is
negligible with respect to the nonlinear localization effect. Forth, we find
the dimer phase, where coupled standing waves are trapped to a few pendulums
at the left-end. Its dynamical origin is the cooperation of the dimerization
and the nonlinear term.

\textit{Discrete sine-Gordon model}. A typical nonlinear system is the
sine-Gordon model described by%
\begin{equation}
m\frac{d^{2}\phi }{dt^{2}}=\kappa \frac{\partial ^{2}\phi }{\partial x^{2}}%
-g\sin \phi .
\end{equation}%
By discretizing it on a one-dimensional chain, we obtain a discrete
sine-Gordon model\cite{HirotaDSG,Orfan}, where the equation of motion is
given by%
\begin{equation}
m\frac{d^{2}\phi _{n}}{dt^{2}}=\kappa \left[ \phi _{n+1}+\phi _{n-1}-2\phi
_{n}\right] -g\sin \phi _{n}.
\end{equation}%
It is rewritten in the form of%
\begin{equation}
m\frac{d^{2}\phi _{n}}{dt^{2}}+g\sin \phi _{n}-\sum_{nm}M_{nm}\phi _{m}=0,
\label{DSG}
\end{equation}%
with the hopping matrix with the coupling $\kappa $,%
\begin{equation}
M_{nm}=\kappa \left( \delta _{n,m+1}+\delta _{n,m-1}-2\delta _{n,m}\right) .
\end{equation}%
Eq.(\ref{DSG}) is derived from the Lagrangian%
\begin{equation}
L=\sum_{n}\left[ m\left( \frac{d\phi _{n}}{dt}\right) ^{2}+g\cos \phi _{n}%
\right] +\sum_{nm}M_{nm}\phi _{n}\phi _{m}.  \label{Lag}
\end{equation}%
The corresponding Hamiltonian is%
\begin{equation}
H=\sum_{n}\left[ m\left( \frac{d\phi _{n}}{dt}\right) ^{2}-g\cos \phi _{n}%
\right] -\sum_{nm}M_{nm}\phi _{n}\phi _{m},
\end{equation}%
which is a conserved energy.

\textit{Dimerized sine-Gordon model}. We investigate the system where the
matrix $M_{nm}$ generates a nontrivial topological structure. Such a matrix
is simply given by the SSH model. The equation of motion is given by Eq.(\ref%
{DSG}) together with the hopping matrix%
\begin{eqnarray}
M_{nm} &=&-\left( \kappa _{A}+\kappa _{B}\right) \delta _{n,m}+\kappa
_{A}\left( \delta _{2n,2m-1}+\delta _{2m,2n-1}\right)  \notag \\
&&+\kappa _{B}\left( \delta _{2n,2m+1}+\delta _{2m,2n+1}\right) .
\label{HoppiMatrix}
\end{eqnarray}%
The Lagrangian of the system is given by Eq.(\ref{Lag}) with (\ref%
{HoppiMatrix}), to which we refer as the dimerized sine-Gordon model.

The explicit equations are given by%
\begin{eqnarray}
m\frac{d^{2}\phi _{2n-1}}{dt^{2}} &=&\kappa _{A}\left( \phi _{2n}-\phi
_{2n-1}\right) +\kappa _{B}\left( \phi _{2n-2}-\phi _{2n-1}\right)  \notag \\
&&-g\sin \phi _{2n-1}, \\
m\frac{d^{2}\phi _{2n}}{dt^{2}} &=&\kappa _{B}\left( \phi _{2n+1}-\phi
_{2n}\right) +\kappa _{A}\left( \phi _{2n-1}-\phi _{2n}\right)  \notag \\
&&-g\sin \phi _{2n}.
\end{eqnarray}%
It is convenient to introduce the coupling strength $\kappa $\ and the
dimerization parameter $\lambda $ by 
\begin{equation}
\kappa _{A}=\kappa \left( 1+\lambda \right) ,\quad \kappa _{B}=\kappa \left(
1-\lambda \right) ,  \label{DimerParam}
\end{equation}%
with $|\lambda |\leq 1$. 
\begin{figure}[t]
\centerline{\includegraphics[width=0.48\textwidth]{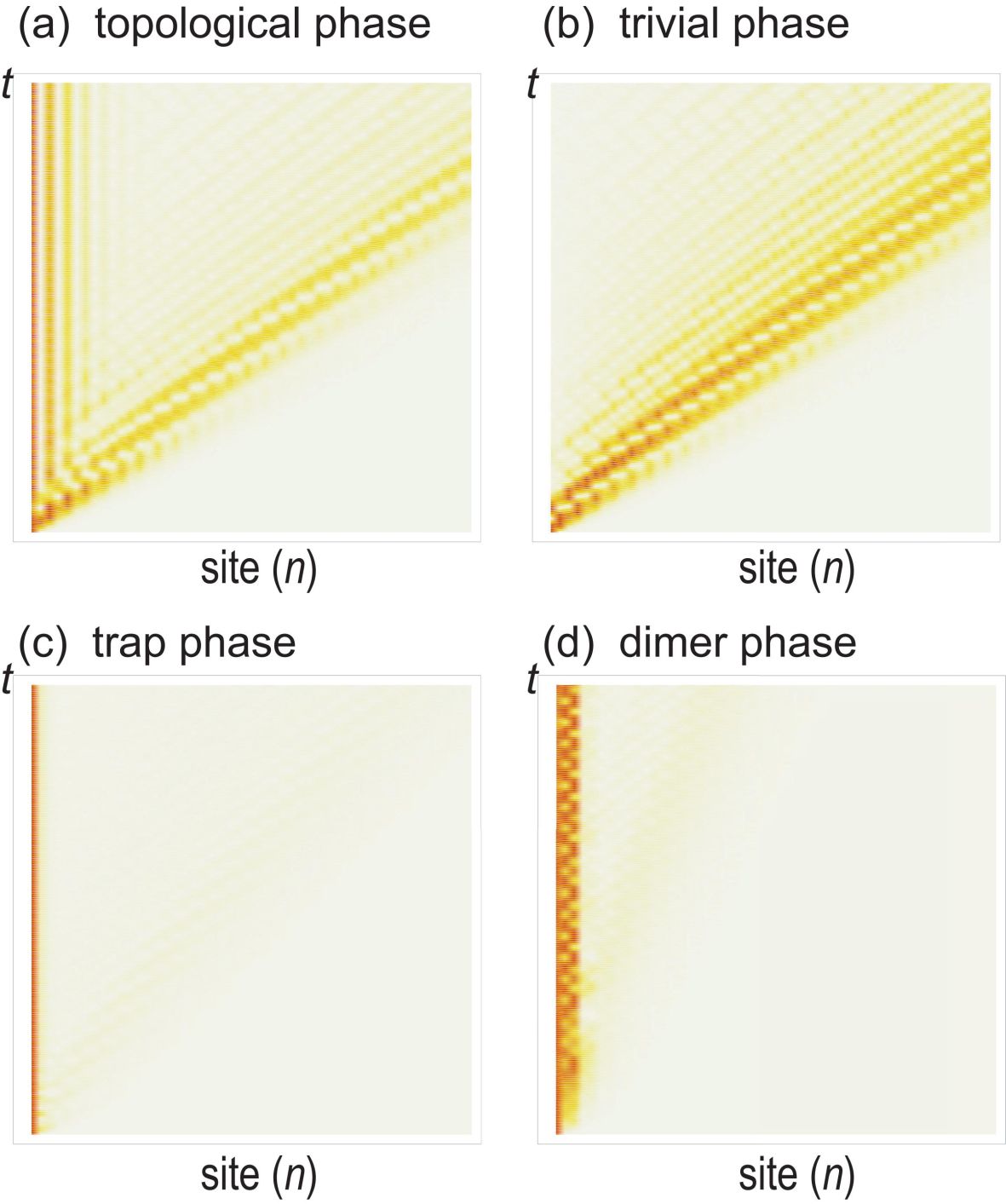}}
\caption{Time evolution of $\protect\phi _{n}$ for (a) the topological phase
with $\protect\lambda =-0.25$ and $\protect\xi =0.1$; (b) the trivial phase
with $\protect\lambda =0.25$and $\protect\xi =0.1$; (c) the trap phase with $%
\protect\lambda =0.5$ and $\protect\xi =0.75$; (d) the dimer phase with $%
\protect\lambda =0.75$ and $\protect\xi =0.5$. We have set $g/\protect\kappa %
=10$.}
\label{FigDynamics}
\end{figure}

\begin{figure*}[t]
\centerline{\includegraphics[width=0.88\textwidth]{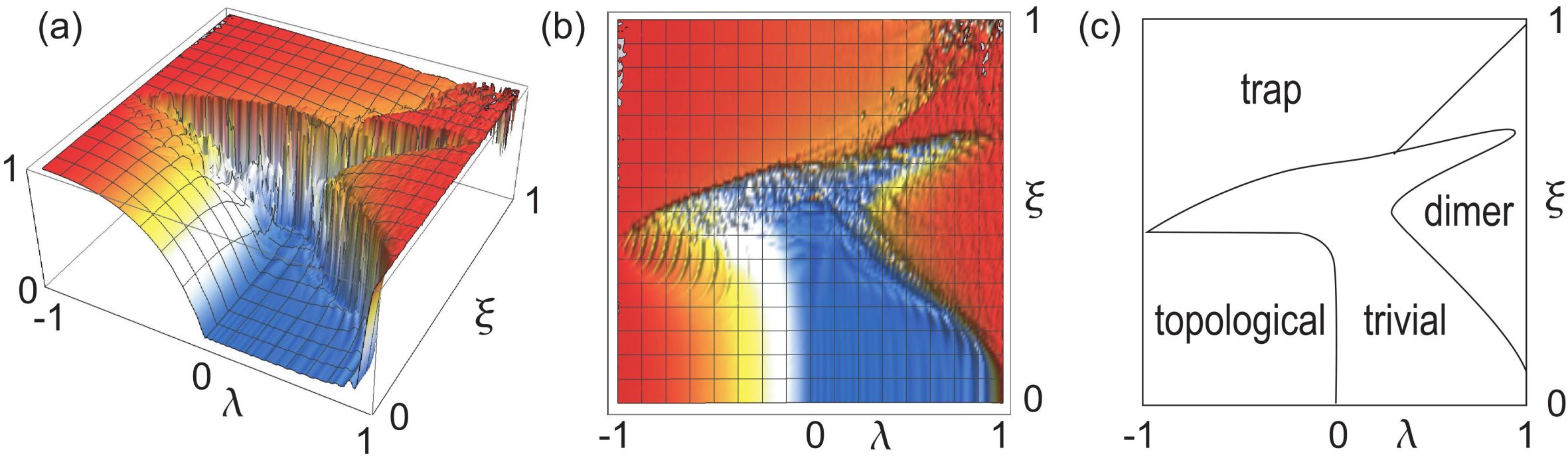}}
\caption{Normalized swing angle $\protect\phi _{1}/\protect\xi \protect\pi $
of the left-end pendulum after enough time as a function of the dimerization 
$\protect\lambda $ and the initial condition $\protect\xi $. (a) Bird's
eye's view and (b) top view. We have set $g/\kappa=10$. (c) Schematic illustration
of phases.}
\label{FigXiPhase}
\end{figure*}

\textit{Mechanical system realization}. We discuss how to realize the
dimerized sine-Gordon model experimentally. It is realized by a mechanical
system shown in Fig.\ref{FigPendulum}, where coupled pendulums are connected
by wires. Each pendulum rotates perpendicular to the wire direction. The
alternating hopping coefficients $\kappa _{A}$ and $\kappa _{B}$ are
introduced by the torsion of the wire connecting two pendulums. $M_{nm}$ is
a matrix representing the couplings between the $n$-th and $m$-th pendulums.
A wire has a restoring force against the force induced by the angle
difference between the two adjacent pendulums. $m$ is an inertia moment of
the pendulum and $g$ is the gravitational acceleration constant.

\textit{Quench dynamics}. We analyze a quench dynamics, where we solve the
dimerized sine-Gordon equation under the initial condition,%
\begin{equation}
\phi _{n}\left( t\right) =\xi \pi \delta _{n,1}\quad \text{and}\quad \dot{%
\phi}_{n}\left( t\right) =0\quad \text{at}\quad t=0,  \label{IniCon}
\end{equation}%
where $|\xi |\leq 1$. Namely, in the case of a chain of pendulums, we set
the initial angle of the left-end pendulum to be $\xi \pi $ and all the
others to be zero as in Fig.\ref{FigPendulum}(a). Then, allowing it to move
with the zero initial velocity under the gravitational force, we study how
the motion propagates along the chain as in Fig.\ref{FigPendulum}(b). We
discuss later that $\xi $\ controls the nonlinearity, and hence we call it
the nonlinearity parameter.

We treat the coupling strength $\kappa $, the dimerization parameter $%
\lambda $ and the nonlinearity parameter $\xi $ as the system parameters. We
have studied the quench dynamics for a variety of these parameters. We have
found there are four types of solutions numerically, whose typical
structures are given in Fig.\ref{FigDynamics}.

In Fig.\ref{FigDynamics}(a), there are standing waves mainly at the left-end
pendulum and weakly at a few adjacent pendulums, and furthermore there are
propagating waves into the bulk with a constant velocity. In Fig.\ref%
{FigDynamics}(b), there are only propagating waves into the bulk with a
constant velocity. In Fig.\ref{FigDynamics}(c), the standing wave is trapped
strictly at the left-end pendulum. In Fig.\ref{FigDynamics}(d), the standing
waves are trapped to a few pendulums at the left-end.

In general, it is impossible to solve the dimerized sine-Gordon equation
analytically. Nevertheless, it is possible to obtain analytical results to
explain these behaviors.

\textit{Topological and trivial phases.} First, we make a change of
variable, $\phi _{n}=\xi \phi _{n}^{\prime }$, and rewrite Eq.(\ref{DSG}) as%
\begin{equation}
m\frac{d^{2}\phi _{n}^{\prime }}{dt^{2}}=\sum_{m}M_{nm}\phi _{m}^{\prime }-%
\frac{g}{\xi }\sin \xi \phi _{n}^{\prime },  \label{Scale}
\end{equation}%
with the initial condition%
\begin{equation}
\phi _{1}^{\prime }(0)=\pi .
\end{equation}%
By taking the limit $\xi \rightarrow 0$, Eq.(\ref{Scale}) is reduced to a
linear equation,%
\begin{equation}
m\frac{d^{2}\phi _{n}}{dt^{2}}=\sum_{m}\overline{M}_{nm}\phi _{m},
\label{WeakDSG}
\end{equation}%
where we have redefined the hopping matrix as%
\begin{equation}
\overline{M}_{nm}\equiv M_{nm}-g\delta _{n,m}.  \label{HoppiMatrixA}
\end{equation}%
Actually, it is a very good approximation to set $\sin \xi \phi _{n}^{\prime
}\simeq \xi \phi _{n}^{\prime }$ in the vicinity of $\xi =0$. We call such a
parameter region the weak nonlinear regime, where the nonlinear equation (%
\ref{DSG}) is well approximated by Eq.(\ref{WeakDSG}). Physically, this
corresponds to the case where the pendulum is approximated by a harmonic
oscillator. Thus, the parameter $\xi $ controls the nonlinearity.

Eq.(\ref{WeakDSG}) is the SSH model with a modified matrix $\overline{M}%
_{nm} $. The SSH model is well known to have a topological and trivial
phases. The topological number is the Berry phase defined by%
\begin{equation}
\Gamma =\frac{1}{2\pi }\int_{9}^{2\pi }A\left( k\right) dk,
\label{ChiralIndexA}
\end{equation}%
where $A\left( k\right) =-i\left\langle \psi (k)\right\vert \partial
_{k}\left\vert \psi (k)\right\rangle $ is the Berry connection with $\psi
(k) $ the eigenfunction of $\overline{M}\left( k\right) $. Note that the
diagonal term in Eq.(\ref{HoppiMatrixA}) with (\ref{HoppiMatrix}) does not
contribute to the topological charge because the wave function $\psi (k)$
does not depend on the diagonal term. Hence, the present model (\ref{WeakDSG}%
) has the same phases as the SSH model. The system is topological for $%
\lambda <0$ and trivial for $\lambda >0$ in the weak nonlinear regime.

The topological phase is characterized by the emergence of zero-mode edge
states for a finite chain. In the linearized equation (\ref{WeakDSG}), the
zero-mode edge state is solved as%
\begin{equation}
\phi _{2n+1}=\left( -\frac{\kappa _{A}}{\kappa _{B}}\right) ^{n}\phi
_{1},\quad \quad \phi _{2n}=0.
\end{equation}%
It has the major component in the left-end pendulum but has components also
in a few adjacent pendulums. Now, the initial motion is given only to the
left-end pendulum, which is only a part of the zero-mode edge state. This
mismatch lets some parts to propagate into the bulk with the velocity $\sim 
\sqrt{\kappa /m}$, while those within the zero-mode edge state stay as they
are, exhibiting standing waves. Thus, this analytic solution well describes
the structure made of standing waves and propagating waves, as shown in Fig.%
\ref{FigDynamics}(a), where $\lambda =-0.25$ and $\xi =0.1$.

On the other hand, there is no zero-mode edge state in the trivial phase.
Hence, the left-end pendulum motion propagates entirely into the bulk, which
well explains the structure made of propagating waves with the velocity $%
\sim \sqrt{\kappa /m}$ in Fig.\ref{FigDynamics}(b), where $\lambda =0.25$
and $\xi =0.1$.

\textit{Trap phase}. We next study the limit where the nonlinear term is
dominant over the hopping term in Eq.(\ref{DSG}), where we may approximate
it as%
\begin{equation}
m\frac{d^{2}\phi _{n}}{dt^{2}}=-g\sin \phi _{n}.  \label{StrongDSG}
\end{equation}%
We call such a parameter region the strong nonlinear regime, where the
nonlinear equation (\ref{DSG}) is well approximated by Eq.(\ref{StrongDSG}).

The prominent feature of the strong nonlinear regime is that all equations
are perfectly decoupled. Each one is a simple pendulum equation, whose exact
solution is given by%
\begin{equation}
\phi _{n}=2\sin ^{-1}[\alpha _{n}\text{sn}(\omega t,\alpha _{n})],
\label{PenduMotionA}
\end{equation}%
where sn is the Jacobi elliptic function, $\omega \equiv \sqrt{g/m}$ and $%
\alpha _{n}$ is determined as $\alpha _{n}=\sin [\phi _{n}\left( 0\right)
/2] $ in terms of the initial condition $\phi _{n}(0)$. Under the initial
condition (\ref{IniCon}), the left-end pendulum makes a motion described by
Eq.(\ref{PenduMotionA}) with $n=1$ while all other pendulums remain
stationary. The pendulum motion is perfectly trapped to the left end of the
chain, as explains the structure made of a single standing wave in Fig.\ref%
{FigDynamics}(c), where $\lambda =0.5$ and $\xi =0.75$.

\textit{Dimer phase}. We finally consider the limit $\lambda =1$, where the
system is perfectly dimerized. In this case, the equations of motion are
given by%
\begin{eqnarray}
m\frac{d^{2}\phi _{1}}{dt^{2}} &=&\kappa _{A}\left( \phi _{2}-\phi
_{1}\right) -g\sin \phi _{1}, \\
m\frac{d^{2}\phi _{2}}{dt^{2}} &=&\kappa _{A}\left( \phi _{1}-\phi
_{2}\right) -g\sin \phi _{2}.
\end{eqnarray}%
By setting $\phi _{2}=-\phi _{1}$, they are combined into one equation%
\begin{equation}
m\frac{d^{2}\phi _{1}}{dt^{2}}=-2\kappa _{A}\phi _{1}-g\sin \phi _{1}.
\end{equation}%
This is solved for $\phi _{1}$ as the inverse function of the integral%
\begin{equation}
\int \frac{d\phi _{1}}{\frac{2}{m}\sqrt{E+g\cos \phi _{1}-\kappa _{A}\phi
_{1}^{2}}}=t-t_{0},
\end{equation}%
where%
\begin{equation}
E=\frac{m}{2}\dot{\phi}_{1}^{2}-g\cos \phi _{1}+\kappa _{A}\phi _{1}^{2}
\end{equation}%
is the conserved energy.

However, the initial condition is given by $\phi _{1}=\xi \pi $\ and $\phi
_{2}=0$, which does not satisfies the condition $\phi _{2}=-\phi _{1}$. It
leads to a complicated behavior at the initial stage. Furthermore, the
above analysis is correct only in the limit $\lambda =1$. In general, a
coupling is present between the dimer and the adjacent pendulum. Numerical
calculation shows coupled standing waves trapped to a few pendulums at the
left-end in Fig.\ref{FigDynamics}(d), where $\lambda =0.75$\ and $\xi =0.5$.
This is a kind of a pull-in effect in nonlinear systems.

\textit{Phase diagram}. We have performed a numerical calculation of the
quench dynamics for the left-end pendulum in a wide range of system
parameters. We show the value of $\phi _{1}$ of the left-end pendulum after
enough time as a function of the dimerization $\lambda $ and the initial
phase $\xi $ in Fig.\ref{FigXiPhase}. It presents a phase diagram of the
system. We have set $g/\kappa =10$ so that the system belongs to the strong
nonlinear regime around $\xi =1$.

We find four distinct phases: 1) the topological phase, 2) the trivial
phase, 3) the trap phase and 4) the dimer phase. The distinction between the
topological and trivial phases is clear analytically in the weak nonlinear
regime ($\xi \simeq 0$). We have found numerically that this topological
phase boundary does not change in spite of the increase of the nonlinear
term up to $\xi \lesssim 0.5$. On the other hand, there is a
nonlinearity-induced trap phase in the strong nonlinear regime ($\xi \simeq
1 $). We have found numerically that the trap phase appear even for $\xi
\gtrsim 0.6$. There is another phase in the vicinity of $\lambda =1$, which
is the dimer phase. We have found numerically that the dimer phase is
realized for $\lambda \gtrsim 0.3$ at $\xi \approx 0.5$.

\textit{Discussion}. We have explored the phase diagram of the dimerized
sine-Gordon model by investigating the quench dynamics. The phase diagram is
very similar to that of the nonlinear Schr\"{o}dinger systems\cite{NLPhoto}
although the models are very different. Indeed, the former is the
second-order differential equation with real variables, while the latter is
the first-order differential equation with complex variables.

The similarity reveals a universal feature of nonlinear topological systems.
The similarity between these models is that there are two competing terms.
One is the hopping term governing topological physics in the weak nonlinear
regime, and the other is the nonlinear term governing nontopological physics
in the strong nonlinear regime. In the weak nonlinear regime, the system is
well described by the hopping term and the topological phase boundary
remains as it is. On the other hand, in the strong nonlinear limit, the
system turns into a nonlinearity-induced trap phase irrespective of the
dimerization parameter $\lambda $, because the hopping term does not play a
significant role. In addition, there is a dimer phase, where the quench
dynamics is trapped to a few pendulums at the left-end.

Our results show that the quench dynamics starting from the edge is a good
signal to determine a phase diagram. It is an interesting problem to study
various nonlinear systems in the context of topology.

The author is very much grateful to N. Nagaosa for helpful discussions on
the subject. This work is supported by the Grants-in-Aid for Scientific
Research from MEXT KAKENHI (Grants No. JP17K05490 and No. JP18H03676). This
work is also supported by CREST, JST (JPMJCR16F1 and JPMJCR20T2).


\begin{thebibliography}{99}
\bibitem{Hasan} M. Z. Hasan and C. L. Kane, Rev. Mod. Phys. \textbf{82},
3045 (2010).

\bibitem{Qi} X.-L. Qi and S.-C. Zhang, Rev. Mod. Phys. \textbf{83}, 1057
(2011).

\bibitem{Prodan} E. Prodan and C. Prodan, Phys. Rev. Lett. \textbf{103},
248101 (2009).

\bibitem{TopoAco} Z. Yang, F. Gao, X. Shi, X. Lin, Z. Gao, Y. Chong and B.
Zhang, Phys. Rev. Lett. \textbf{114}, 114301 (2015).

\bibitem{Berto} P. Wang, L. Lu and K. Bertoldi, Phys. Rev. Lett. \textbf{115}%
, 104302 (2015).

\bibitem{Xiao} M. Xiao, G. Ma, Z. Yang, P. Sheng, Z. Q. Zhang and C. T.
Chan, Nat. Phys. \textbf{11}, 240 (2015).

\bibitem{He} C. He, X. Ni, H. Ge, X.-C. Sun,Y.-B. Chen1 M.-H. Lu, X.-P. Liu,
L. Feng and Y.-F. Chen, Nature Physics \textbf{12}, 1124 (2016).

\bibitem{Abba} H. Abbaszadeh, A. Souslov, J. Paulose, H. Schomerus and V.
Vitelli, Phys. Rev. Lett. \textbf{119}, 195502 (2017).

\bibitem{Xue} H. Xue, Y. Yang, F. Gao, Y. Chong and B.Zhang, Nature
Materials \textbf{18}, 108 (2019).

\bibitem{Ni} X. Ni, M. Weiner, A. Alu and A. B. Khanikaev, Nature Materials 
\textbf{18}, 113 (2019).

\bibitem{Wei} M. Weiner, X. Ni, M. Li, A. Alu, A. B. Khanikaev, Science
Advances \textbf{6}, eaay4166 (2020).

\bibitem{Xue2} H. Xue, Y. Yang, G. Liu, F. Gao, Y. Chong and B. Zhang, Phys.
Rev. Lett. \textbf{122}, 244301 (2019).

\bibitem{Lubensky} C. L. Kane and T. C. Lubensky, Nature Phys. \textbf{10},
39 (2014).

\bibitem{Chen} B. Gin-ge Chen, N. Upadhyaya and V. Vitelli, PNAS \textbf{111}%
, 13004 (2014).

\bibitem{Nash} L. M. Nash, D. Kleckner, A. Read, V. Vitelli, A. M. Turner
and W. T. M. Irvine, PNAS \textbf{112}, 14495 (2015).

\bibitem{Paul} J. Paulose, A. S. Meeussen and V. Vitelli, PNAS \textbf{112},
7639 (2015).

\bibitem{Sus} R. Susstrunk, S. D. Huber, Science \textbf{349}, 47 (2015).

\bibitem{Sss} R. Susstrunk and S. D. Huber, Proc. Natl. Acad. Sci. USA 
\textbf{113}, E4767 (2016).

\bibitem{Huber} S. D. Huber, Nature Physics \textbf{12}, 621 (2016).

\bibitem{Mee} A. S. Meeussen, J. Paulose and V. Vitelli, Phys. Rev. X 
\textbf{6}, 041029 (2016).

\bibitem{Kariyado} T. Kariyado and Y. Hatsugai, Sci. Rep. \textbf{5}, 18107
(2016).

\bibitem{Hannay} T. Kariyado and Y. Hatsugai, J. Phys. Soc. Jpn. \textbf{85}%
, 043001 (2016).

\bibitem{Po} H. C. Po, Y. Bahri and A. Vishwanath, Phys. Rev. B \textbf{93},
205158 (2016).

\bibitem{Rock} D. Zeb Rocklin, Bryan Gin--ge Chen, Martin Falk, Vincenzo
Vitelli, and T.\thinspace C. Lubensky, Phys. Rev. Lett. \textbf{116}, 135503
(2016).

\bibitem{Takahashi} Y. Takahashi, T. Kariyado and Y. Hatsugai, New J. Phys. 
\textbf{19}, 035003 (2017).

\bibitem{Mat} K. H. Matlack, M. Serra-Garcia, A. Palermo, S. D. Huber and C.
Daraio, Nature Mat. \textbf{17}, 323 (2018).

\bibitem{Taka} Y. Takahashi, T. Kariyado and Y. Hatsugai, Phys. Rev. B 
\textbf{99}, 024102 (2019).

\bibitem{Ghatak} A. Ghatak, M. Brandenbourger, J. van Wezel and C. Coulais,
Proc. Natl. Ac. Sc. U.S.A. \textbf{117}, 29561 (2020).

\bibitem{Wakao} H. Wakao, T. Yoshida, H. Araki, T. Mizoguchi and Y.
Hatsugai, Phys. Rev. B \textbf{101}, 094107 (2020).

\bibitem{KhaniPhoto} A. B. Khanikaev, S. H. Mousavi, W.-K. Tse, M.
Kargarian, A. H. MacDonald, G. Shvets, Nature Materials \textbf{12}, 233
(2013).

\bibitem{Hafe2} M. Hafezi, E. Demler, M. Lukin, J. Taylor, Nature Physics 
\textbf{7}, 907 (2011).

\bibitem{Hafezi} M. Hafezi, S. Mittal, J. Fan, A. Migdall, J. Taylor, Nature
Photonics \textbf{7}, 1001 (2013).

\bibitem{WuHu} L.H. Wu and X. Hu, Phys. Rev. Lett. \textbf{114}, 223901
(2015).

\bibitem{TopoPhoto} L. Lu. J. D. Joannopoulos and M. Soljacic, Nature
Photonics \textbf{8}, 821 (2014).

\bibitem{Ozawa} T. Ozawa, H. M. Price, A. Amo, N. Goldman, M. Hafezi, L. Lu,
M. C. Rechtsman, D. Schuster, J. Simon, O. Zilberberg and L. Carusotto, Rev.
Mod. Phys. \textbf{91}, 015006 (2019).

\bibitem{Hassan} A. E. Hassan, F. K. Kunst, A. Moritz, G. Andler, E. J.
Bergholtz, M. Bourennane, Nature Photonics \textbf{13}, 697 (2019).

\bibitem{Li} M. Li, D. Zhirihin, D. Filonov, X. Ni, A. Slobozhanyuk, A. Alu
and A. B. Khanikaev, Nature Photonics \textbf{14}, 89 (2020).

\bibitem{TECNature} S. Imhof, C. Berger, F. Bayer, J. Brehm, L. Molenkamp,
T. Kiessling, F. Schindler, C. H. Lee, M. Greiter, T. Neupert, R. Thomale,
Nat. Phys. \textbf{14}, 925 (2018).

\bibitem{ComPhys} C. H. Lee , S. Imhof, C. Berger, F. Bayer, J. Brehm, L. W.
Molenkamp, T. Kiessling and R. Thomale, Communications Physics, \textbf{1},
39 (2018).

\bibitem{Hel} T. Helbig, T. Hofmann, C. H. Lee, R. Thomale, S. Imhof, L. W.
Molenkamp and T. Kiessling, Phys. Rev. B \textbf{99}, 161114 (2019).

\bibitem{Lu} Y. Lu, N. Jia, L. Su, C. Owens, G. Juzeliunas, D. I. Schuster
and J. Simon, Phys. Rev. B \textbf{99}, 020302 (2019).

\bibitem{YLi} Y. Li, Y. Sun, W. Zhu, Z. Guo, J. Jiang, T. Kariyado, H. Chen
and X. Hu, Nat. Com. \textbf{9}, 4598 (2018).

\bibitem{EzawaTEC} M. Ezawa, Phys. Rev. B \textbf{98}, 201402(R) (2018).

\bibitem{Zhao} E. Zhao, Ann. Phys. \textbf{399}, 289 (2018).

\bibitem{EzawaLCR} M. Ezawa, Phys. Rev. B \textbf{99}, 201411(R) (2019).

\bibitem{EzawaSkin} M. Ezawa, Phys. Rev. B \textbf{99}, 121411(R) (2019).

\bibitem{Garcia} M. Serra-Garcia, R. Susstrunk and S. D. Huber, Phys. Rev. B 
\textbf{99}, 020304 (2019).

\bibitem{Hofmann} T. Hofmann, T. Helbig, C. H. Lee, M. Greiter, R. Thomale,
Phys. Rev. Lett. \textbf{122}, 247702 (2019).

\bibitem{Snee} D. D. J. M. Snee, Y.-P. Ma, Extreme Mechanics Letters 100487
(2019).

\bibitem{PWLo} P.-W. Lo, K. Roychowdhury, B. G.-g. Chen, C. D. Santangelo,
C.-M. Jian, M. J. Lawler, Phys. Rev. Lett. \textbf{127}, 076802 (2021).

\bibitem{MechaRot} M. Ezawa, arXiv:2108.09634.

\bibitem{MacZ} Lukas J. Maczewsky, Matthias Heinrich, Mark Kremer, Sergey K.
Ivanov, Max Ehrhardt, Franklin Martinez, Yaroslav V. Kartashov, Vladimir V.
Konotop, Lluis Torner, Dieter Bauer, Alexander Szameit, Science \textbf{370}%
, 701 (2010).

\bibitem{Ley} D. Leykam and Y. D. Chong, Phys. Rev. Lett. \textbf{117},
143901 (2016).

\bibitem{Zhou} X. Zhou, Y. Wang, D. Leykam and Y. D. Chong, New J. Phys. 
\textbf{19}, 095002 (2017).

\bibitem{Kruk} S. Kruk, A. Poddubny, D. Smirnova, L. Wang, A. Slobozhanyuk,
A. Shorokhov, I. Kravchenko, B. Luther-Davies and Y. Kivshar, Nature
Nanotechnology \textbf{14}, 126 (2019).

\bibitem{Smi} D. Smirnova, D. Leykam, Y. Chong and Y. Kivshar, Applied
Physics Reviews \textbf{7}, 021306 (2020).

\bibitem{Tulo} T. Tuloup, R. W. Bomantara, C. H. Lee and J. Gong, Phys. Rev.
B \textbf{102}, 115411 (2020).

\bibitem{NLPhoto} M. Ezawa, arXiv:2110.06578.

\bibitem{Kirch} M. S. Kirsch, Y. Zhang, M. Kremer, L. J. Maczewsky, S. K.
Ivanov, Y. V. Kartashov, L. Torner, D. Bauer, A. Szameit and M. Heinrich,
Nature Physics \textbf{17}, 995 (2021).

\bibitem{Hadad} Y. Hadad, J. C. Soric, A. B. Khanikaev, and A. Al\`{u},
Nature Electronics \textbf{1}, 178 (2018).

\bibitem{TopoToda} M. Ezawa, cond-mat/arXiv:2105.10851.

\bibitem{Zange} F. Zangeneh-Nejad and R. Fleury, Phys. Rev. Lett. \textbf{123%
}, 053902 (2019).

\bibitem{HirotaDSG} R. Hirota, J. Phys. Soc. Jpn, \textbf{43}, 2079 (1977).

\bibitem{Orfan} S. J. Orfanidis, Phys. Rev. B \textbf{18}, 3822 (1978).

\end{thebibliography}
\end{document}